\documentclass[showpacs,twocolumn,prl,superscriptaddress]{revtex4}
\usepackage{amsfonts}
\usepackage{amsmath}
\usepackage{amssymb}
\usepackage{graphicx}
\usepackage{verbatim}
\usepackage{textcomp}

\providecommand{\U}[1]{\protect\rule{.1in}{.1in}}

\begin{document}

\title{Ultra-broadband and sensitive cavity optomechanical magnetometry}

\author{Bei-Bei Li}
\affiliation{Institute of Physics, Chinese Academy of Sciences, Beijing 100190, P. R. China.}

\author{George Brawley}
\affiliation{School of Mathematics and Physics, The University of Queensland, St Lucia, Queensland 4072, Australia.}

\author{Hamish Greenall}
\affiliation{School of Mathematics and Physics, The University of Queensland, St Lucia, Queensland 4072, Australia.}

\author{Stefan Forstner}
\affiliation{School of Mathematics and Physics, The University of Queensland, St Lucia, Queensland 4072, Australia.}

\author{Eoin Sheridan}
\affiliation{School of Mathematics and Physics, The University of Queensland, St Lucia, Queensland 4072, Australia.}

\author{Halina Rubinsztein-Dunlop}
\affiliation{School of Mathematics and Physics, The University of Queensland, St Lucia, Queensland 4072, Australia.}

\author{Warwick P. Bowen}
\email{w.bowen@uq.edu.au}
\affiliation{School of Mathematics and Physics, The University of Queensland, St Lucia, Queensland 4072, Australia.}





\date{\today}

\begin{abstract}

Magnetostrictive optomechanical cavities provide a new optically-readout approach to room temperature magnetometry. Here we report ultrasensitive and ultrahigh bandwidth cavity optomechanical magnetometers constructed by embedding a grain of the magnetostrictive material Terfenol-D within a high quality ($Q$) optical microcavity on a silicon chip. By engineering their physical structure, we achieve a peak sensitivity of 26~pT/$\sqrt{\mathrm{Hz}}$ comparable to the best cryogenic microscale magnetometers, along with a 3~dB bandwidth as high as 11.3~MHz. Two classes of magnetic response are observed, which we postulate arise from the crystallinity of the Terfenol-D. This allows single- and poly-crystalline grains to be distinguished at the level of a single particle. Our results may enable applications such as lab-on-chip nuclear magnetic spectroscopy and magnetic navigation.

\end{abstract}

\maketitle

\section{Introduction}

The resonant enhancement of both optical and mechanical response in a cavity optomechanical system \cite{2008Science,2014RMP} has enabled precision sensors \cite{2014APR} of displacement\cite{2008NJP,GW}, force \cite{force}, mass \cite{mass}, acceleration \cite{acceleration1,acceleration2}, ultrasound \cite{ultrasound}, and magnetic fields \cite{magnetic1,magnetic2,magnetic3,magnetic4,magnetic5,magnetic6}. Cavity optomechanical magnetometers are particulary attractive, promising state-of-the-art sensitivity without the need for cryogenics, with only microwatt power consumption \cite{magnetic1,magnetic2,magnetic3,magnetic4,magnetic6}, and with silicon chip based fabrication offering scalability \cite{magnetic5}.
For instance, cavity optomechanical magnetometers working in the megahertz frequency range have been demonstrated by using a magnetostrictive material Terfenol-D, either manually deposited onto a microcavity \cite{magnetic1,magnetic2,magnetic4} with a reported peak sensitivity of 200~pT/$\sqrt{\mathrm{Hz}}$ \cite{magnetic2}, or sputter coated onto the microcavity with a reported peak sensitivity of 585~pT/$\sqrt{\mathrm{Hz}}$ \cite{magnetic5}. Efforts have also been made to improve the sensitivity at the hertz-to-kilohertz frequency range, which is relevant to many applications. The nonlinearity inherent in magnetostrictive materials have been used to mix the low-frequency signals up to high-frequency \cite{magnetic2}; a peak sensitivity of 131~pT/$\sqrt{\mathrm{Hz}}$ has been reported in the 100~kHz range using a centimeter-sized CaF$_{2}$ cavity with a cylinder of Terfenol-D crystal embedded inside \cite{magnetic3}; and polymer coated microcavities have been combined with millimeter-sized magnets to provide a sensitivity of 880~pT/$\sqrt{\mathrm{Hz}}$ at 200~Hz \cite{magnetic6}. Resonant magnon assisted optomechanical magnetometers have recently been realized, achieving a sensitivity of 103~pT/$\sqrt{\mathrm{Hz}}$ in the GHz frequency range \cite{GHz}, while torque magnetometers have also been demonstrated using nanomechanical systems for magnetization measurement \cite{Torque1,Torque2}. With all of this recent progress, however, the sensitivity of the best optomechanical magnetometers remains around an order-of-magnitude inferior to similarly-sized cryogenic magnetometers \cite{SQUID1,SQUID2,SQUID3}.


This paper focuses on optimizing the sensitivity of magnetostrictive optomechanical magnetometers. We find that the sensitivity depends critically on the shape of the support that suspends the magnetometer above the silicon chip. This support both constrains the magnetostriction-induced mechanical motion and provides an avenue for thermal fluctuations to enter the system. By engineering its structure to increase both its compliance and the mechanical quality factor of the device, we demonstrate around an order of magnitude improvement in sensitivity compared to previous works, to 26~pT/$\sqrt{\mathrm{Hz}}$. This is comparable to the similarly-sized cryogenic magnetometers \cite{SQUID1,SQUID2,SQUID3}.




The magnetic response as a function of magnetic field frequency is found to show two significantly different behaviors: a relatively smooth response modulated by the mechanical resonances of the structure; and a response that exhibits dramatic variations as a function of frequency, with these variations occurring under the envelope of the mechanical resonances. We refer to these two behaviors as Type~I and Type~II, respectively. The magnetic response of the Type~II devices is observed to be highly sensitive to direct current (DC) magnetic fields. We find this behavior is consistent with interference of acoustic waves produced at multiple grain boundaries in a polycrystalline Terfenol-D particle. We therefore infer that the Type~I and II responses arise when the particle is mono- and poly-crystalline, respectively. Our devices, therefore, provide a method to characterize the Terfenol-D crystal structure, a measurement which is generally challenging at the level of a single grain.


The magnetometers show an ultra-broadband response. The working frequency ranges for both Type~I and Type~II magnetometers are more than 130~MHz, limited by the bandwidth of the photoreceivers we use in our experiment. Accumulated 3~dB bandwidths \cite{magnetic4} of 11.3~MHz and 120~kHz are measured for the Type~I and Type~II magnetometers, respectively. This compares favorably to other sensitive optically-read-out magnetometers, which typically have bandwidths in the 1-10~kHz range \cite{atomic1,atomic2,NV1,NV2}.


\section{Experiment and result}
\subsection{Fabrication}

\begin{figure}[t!]
\centering
\includegraphics[width=8cm]{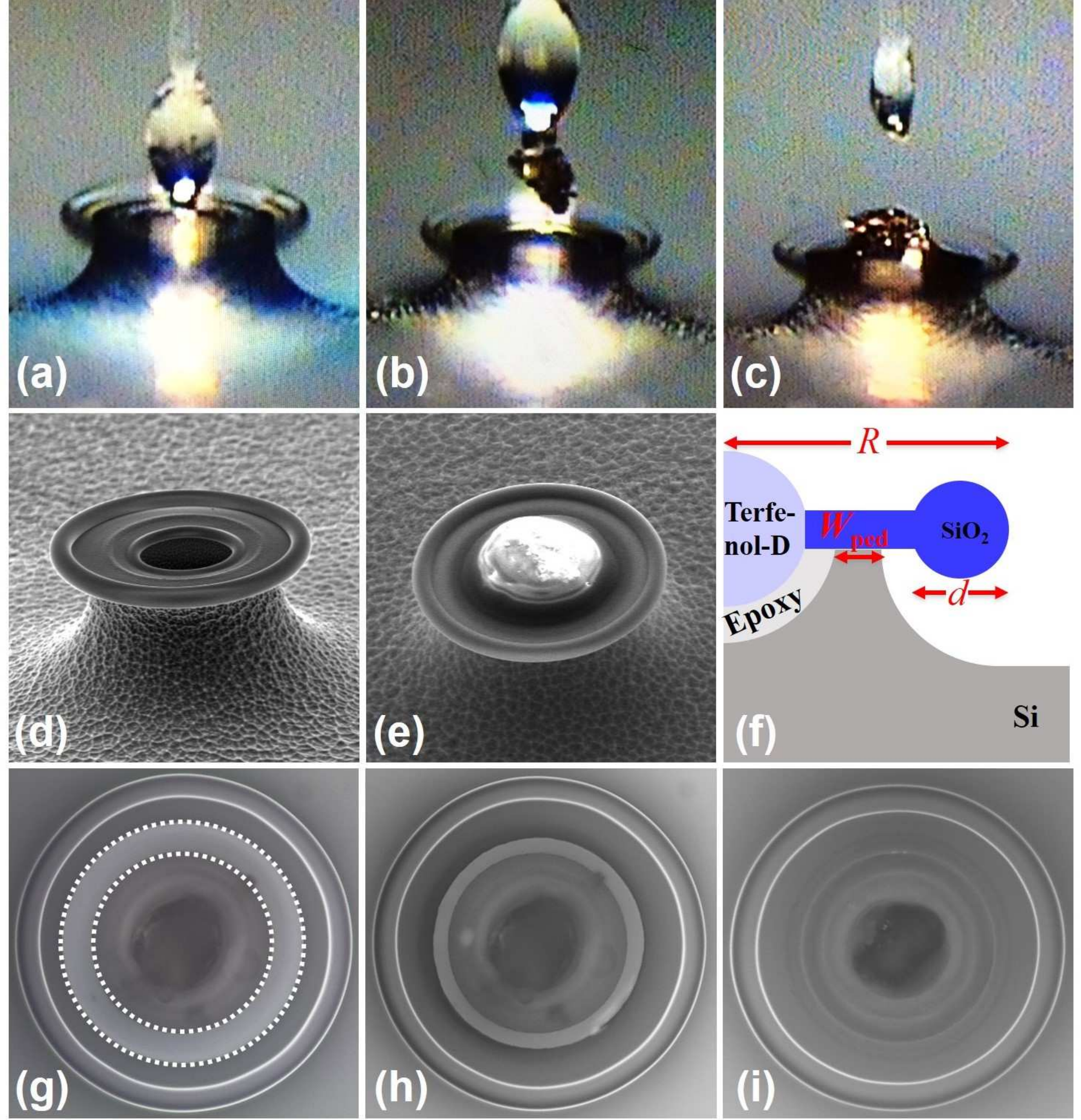}
\caption{(Color online). (a)-(c), Optical microscope images showing the Terfenol-D deposition process. (d)-(e), The SEM images of a microtoroid before and after the Terfenol-D deposition. (f), A schematic of the side-view of a magnetometer, with a principal radius of $R$, minor diameter of $d$, and a pedestal width of $W_\textrm{ped}$. (g)-(i), Top-view optical microscope images of a fabricated magnetometer, with gradually decreased pedestal width, marked in the area between the two white dotted circles.}
\label{fig1}
\end{figure}

The magnetometers are fabricated by depositing Terfenol-D particles into holes etched into the center of silica microtoroids, following the approach in Ref. \cite{magnetic2}, as shown in Figs. 1(a)-1(c). The silica microtoriods with central holes are first fabricated through photolithography, hydrofluoric acid etching, xenon difluoride (XeF$_{2}$) etching, and carbon dioxide (CO$_{2}$) laser reflow process \cite{toroid}. We then use a fiber tip to deposit a droplet of epoxy into the hole (Fig. 1(a)), and use the same fiber tip to pick up a piece of Terfenol-D and place it into the epoxy inside the hole (Figs. 1(b)-(c)). The epoxy is then cured over a period of 8~hours, to provide the bonded magnetometer. Figures 1(d) and 1(e) are scanning electron microscope (SEM) images of one silica microtoroid before and after the Terfenol-D deposition. Figure 1(f) shows a side-view schematic of the magnetometer, with a principal radius of $R$, minor diameter of $d$, and a pedestal width of $W_{\mathrm{ped}}$.

\subsection{Measurement of magnetic field sensitivity}

\begin{figure*}[t!]
\centering
\includegraphics[width=12cm]{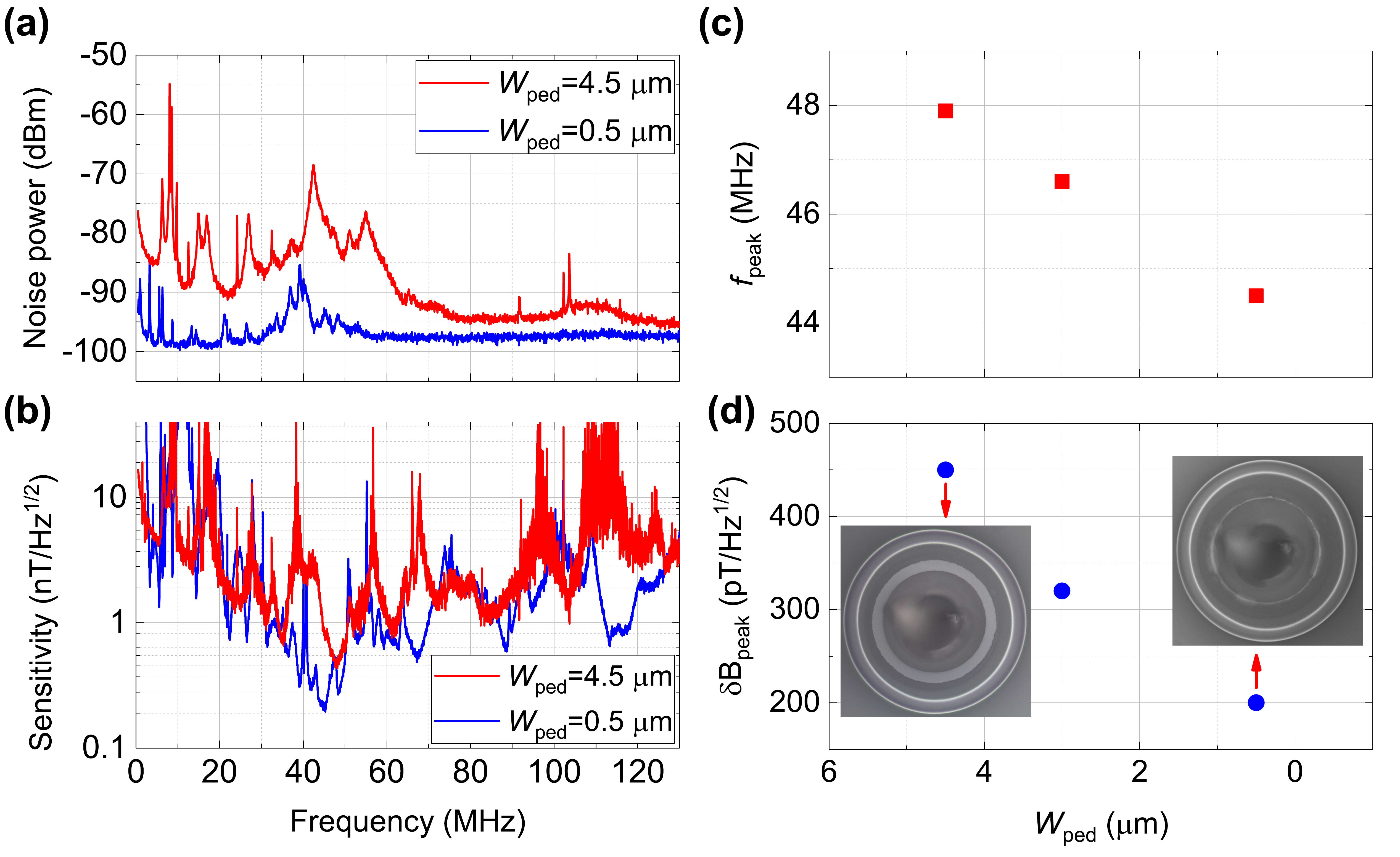}
\caption{(Color online). Magnetic field sensitivity improvement by etching down the width of the silicon pedestal. (a)-(b), The noise power spectrum and sensitivity spectra for a magnetometer with pedestal width of 4.5~\textmu m (red curve) and 0.5~\textmu m (black curve). (c)-(d), The peak sensitivity frequency (c) and peak sensitivity (d) of the magnetometer, as a function of the pedestal width.}
\label{fig2}
\end{figure*}

\begin{figure*}[t!]
\centering
\includegraphics[width=12cm]{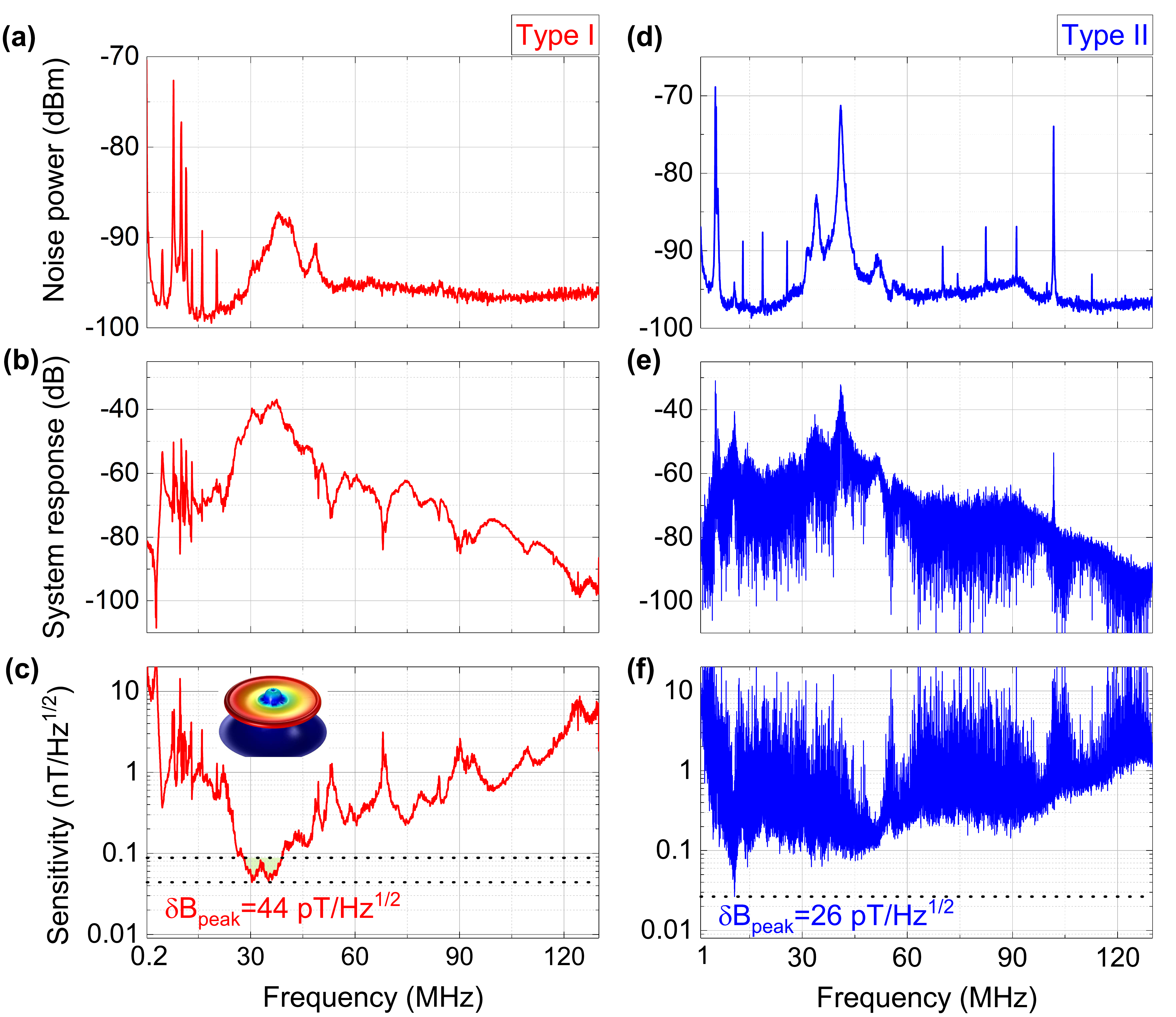}
\caption{(Color online). Measurement results for the two types of magnetometers. (a)-(c), The noise power spectrum, system response, and the sensitivity spectrum for a Type~I magnetometer. The inset of Fig. 3(c) shows the profile of the radial breathing mode (where the peak sensitivity occurs) of the magnetometer, obtained through finite element method simulation using COMSOL Multiphysics. (d)-(f), The corresponding results for a Type~II magnetometer. The peak sensitivities are 44~pT/$\sqrt{\mathrm{Hz}}$ and 26~pT/$\sqrt{\mathrm{Hz}}$ for the Type~I magnetometer and the Type~II magnetometer, respectively.}
\label{fig3}
\end{figure*}

To measure the magnetic field sensitivity of the fabricated magentometers, we use a tapered fiber \cite{taper} to couple light from a tunable laser in the 1550~nm wavelength band into one of their whispering gallery modes (WGMs). We then use a photoreceiver to detect the light transmitted from the microtoroids back into the tapered optical fiber. After we identify a high-$Q$ WGM, we thermally lock the laser frequency on the blue-side of the mode \cite{locking}. Mechanical motion due to the applied magnetic field modulates the perimeter of the device and therefore changes the optical resonance. This translates into a periodical modulation in the transmitted light intensity. We use a spectrum analyzer (SA) to measure the noise power spectrum. We then apply a magnetic field to the magnetometer using a coil driven by a network analyzer. This allows the frequency of the magnetic field applied to the magnetometer to be swept and the magnetic response at each frequency to be characterized. With the noise power spectra and system response, we derive the magnetic field sensitivity, following Ref. \cite{magnetic1}.

\subsection{Sensitivity improvement by silicon pedestal etching}

In our experiment, to achieve a uniform CO$_{2}$ laser reflow process, the silicon pedestal is left with a width of $\sim$5-10~\textmu m after the XeF$_{2}$ etching. Here, in order to improve the mechanical compliance and thus improve the magnetic field sensitivity, we then further etch down the silicon pedestal by performing several runs of XeF$_{2}$ etching after the Terfenol-D deposition process is complete. The width of the pedestal can be directly measured from an optical microscope image, and is marked in the area between the two white dashed circles in Fig. 1(g). Figures 1(g)-(i) are optical microscope images of a magnetometer with gradually decreased pedestal width.

We etch down the silicon pedestal by a few \textmu m in each run of XeF$_{2}$ etching, and measure the sensitivity. In Figs. 2(a) and 2(b), we plot the noise power spectra and sensitivity of a magnetometer with $W_\textrm{ped}$=4.5~\textmu m (red curves), and $W_\textrm{ped}$=0.5~\textmu m (blue curves), respectively. We find that, as $W_{\mathrm{ped}}$ decreases, the mechanical resonances move to lower frequency, due to the increased mechanical compliance and therefore decreased spring constant of the microtoroids. The sensitivity improves across almost the entire active frequency range. In Figs. 2(c) and 2(d), we plot respectively the frequency of the peak sensitivity and the peak sensitivity, as a function of the pedestal width. The inset of Fig. 2(d) shows optical microscope images of the magnetometers with pedestal widths of 4.5~\textmu m (left) and 0.5~\textmu m (right), respectively. It can be seen that, when the pedestal width is decreased from 4.5~\textmu m to 0.5~\textmu m, the peak sensitivity of the magnetometer is improved from 450~pT/$\sqrt{\mathrm{Hz}}$ at 47.7~MHz to 200~pT/$\sqrt{\mathrm{Hz}}$ at 44.5~MHz. This sensitivity improvement from silicon pedestal etching is consistently observed in most magnetometers.

\subsection{Peak sensitivity}

\begin{figure}[b!]
\centering
\centering
\includegraphics[width=7cm]{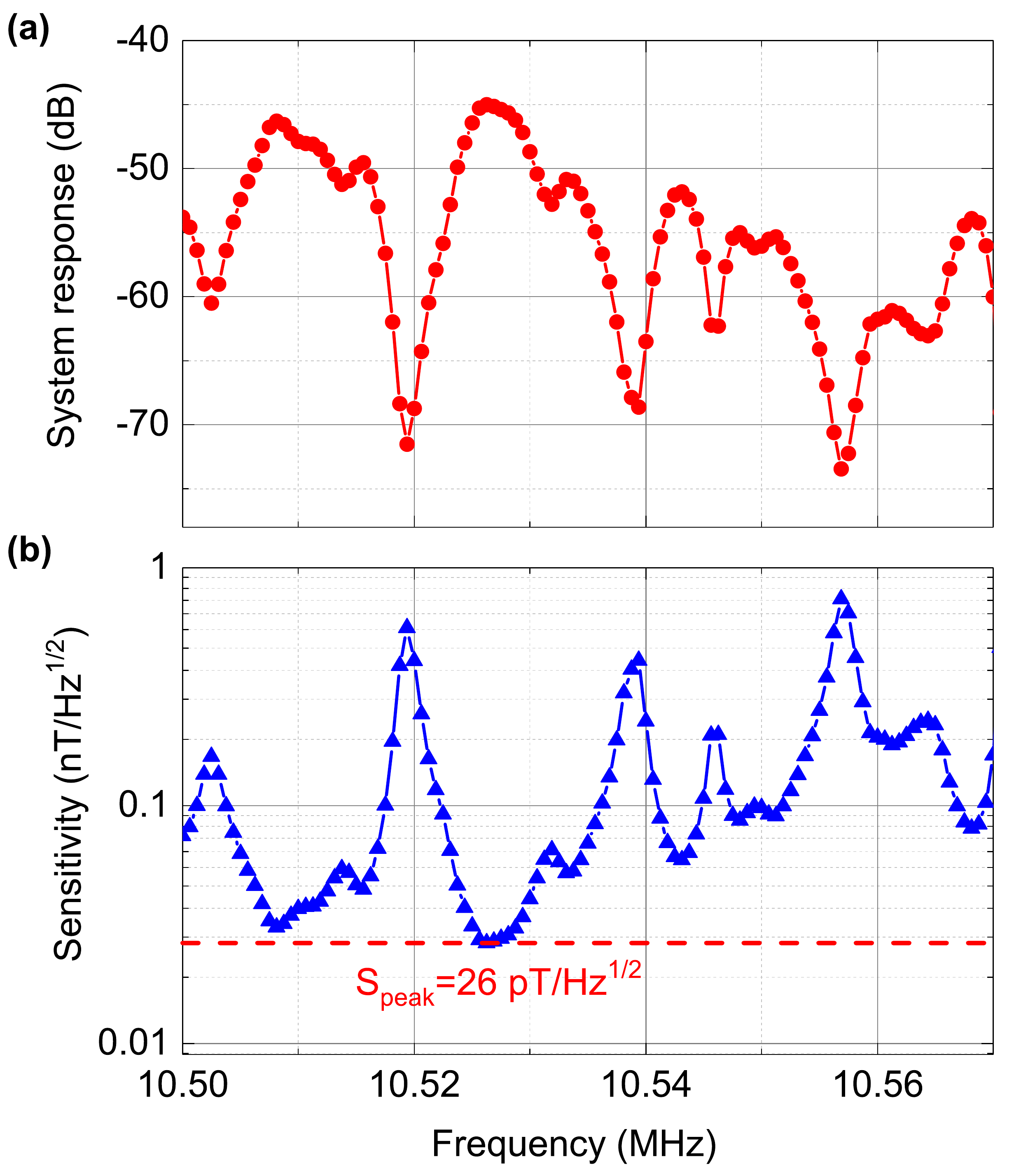}
\caption{(Color online) Zoom-in on the system response (a) and the sensitivity spectrum (b) of the Type~II magnetometer around its peak sensitivity frequency. The peak sensitivity is around 26~pT/$\sqrt{\mathrm{Hz}}$.}
\label{fig4}
\end{figure}

In Fig. 3 we plot the noise power spectra, system responses, and sensitivity spectra of two magnetometers with pedestal widths less than 5 \textmu m. Characterizing several magnetometers, we observed that they each exhibit one of the two distinct behaviors, as illustrated on the left and right columns of Fig. 3. We define magnetometers that fit into the two classes as Type~I and Type~II, respectively. For the Type~I magnetometer, the magnetic response is relatively smooth as a function of the frequency, with peaks corresponding to the mechanical resonances. Figs. 3(a) and 3(b) show the noise power spectrum and the system response of a Type~I magnetometer. For Type~II magnetometers, in addition to the mechanical resonance peaks, there exists strong modulation of magnetic response as a function of the frequency, as shown in Fig. 3(d). This phenomenon will be further studied in the following section.

The peak sensitivity found for Type~I magnetometers is $\sim$44~pT/$\sqrt{\mathrm{Hz}}$ at about 30~MHz. This frequency corresponds to the radial breathing mode of the magnetometer, with its mode profile shown in the inset of Fig. 3(e)), obtained through finite element method simulation using COMSOL Multiphysics. Generally this mode has the largest spatial overlap with the magnetostriction of the Terfenol-D. For the Type~II magnetometer, both the magnetic response and sensitivity vary significantly with small changes in the frequency of the magnetic field, as shown in both Figs. 3(e) and 3(f) and in their zoom-ins shown in Figs. 4(a)-4(b). It can be seen that the magnetic response varies by more than 25~dB within a frequency range of 20~kHz. The peak sensitivity is $\sim$26~pT/$\sqrt{\mathrm{Hz}}$, at $\sim$10.5~MHz, and is flat over a 74~kHz frequency band, which is comparable with other high precision magnetometers, such as atomic magnetometers \cite{atomic1,atomic2} and nitrogen vacancy center magnetometers \cite{NV1,NV2}. This sensitivity is state-of-the-art in micro-optomechanical systems, and is comparable to the sensitivity of micro-sized SQUIDs \cite{SQUID1,SQUID2,SQUID3}.

\subsection{Magnetic response}

\begin{figure}[htb]
\centering
\includegraphics[width=8cm]{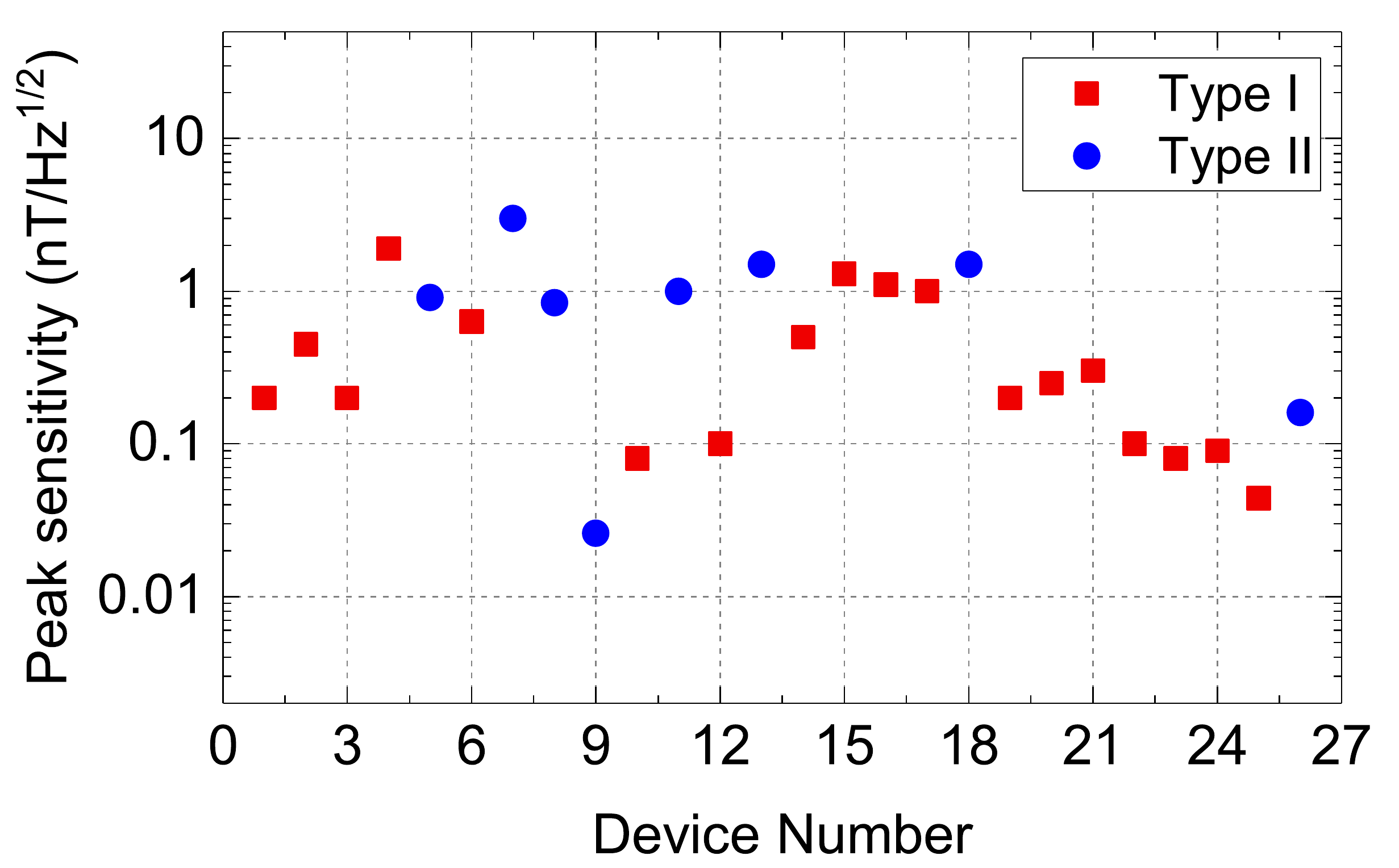}
\caption{(Color online) Measured peak sensitivities of 26 magnetometers, in which 18 of them show Type~I magnetic response (red squares) and 8 show Type~II magnetic response (blue circles).}
\label{fig5}
\end{figure}

In order to study the two types of magnetic response, we fabricated a total of 26 magnetometers. Of them we found that 18 exhibited Type~I magnetic response, and 8 showed Type~II magnetic response. In Fig. 4 we plot the peak sensitivities of the 26 magnetometers, with the red squares denoting the Type~I magnetometers, and the blue circles the Type~II magnetometers. The results in Fig. 2 correspond to device No. 25 (Type~I) and No. 9 (Type~II), respectively. We note, also, that the Type~I magnetometers have a more reliable performance, with five of them having sensitivity better than 100~pT/$\sqrt{\mathrm{Hz}}$. The average sensitivities of the Type~I and Type~II magnetometers are 470~pT/$\sqrt{\mathrm{Hz}}$ and 1.1~nT/$\sqrt{\mathrm{Hz}}$, respectively.

\begin{figure}[htb]
\centering
\includegraphics[width=7.5cm]{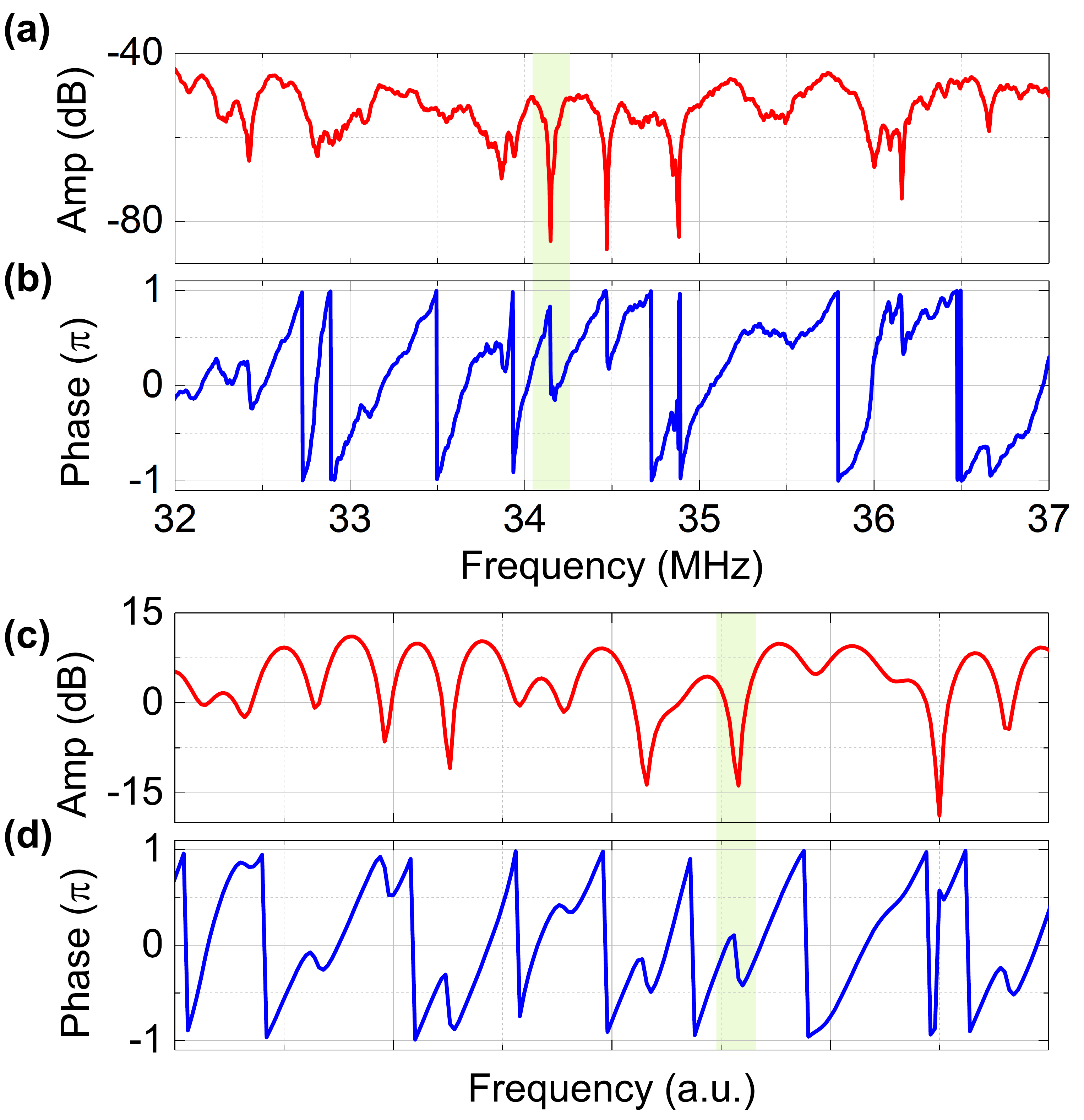}
\caption{(Color online). (a)-(b), Measured amplitude and phase of the magnetic response in the frequency range between 32-37 MHz, for a Type~II magnetometer (No. 26 in Fig. \ref{fig4}). (c) and (d), Theoretically generated amplitude and phase of the system response obtained from the interference of multiple waves from different sources with different amplitudes and phases.}
\label{fig6}
\end{figure}

All of the measured magnetometers were fabricated using the same method. The fact that magnetic domains in Terfenol-D can have a size similar to size of the grains \cite{handbook,microstructure1,microstructure2} used in our experiment suggests that the two different types of magnetic responses originate from the variations in the number and configurations of the crystal grains. In particular, we postulate that the Terfenol-D particles we used in the Type~I and II magnetometers are single crystalline and poly-crystalline, respectively. Since the magnetostriction of Terfenol-D is caused by the domain boundary movement in the presence of a DC magnetic field, this would mean that the Type~II response could be engineered by an external magnetic field. To confirm this, we then take one Type~II magnetometer (device No. 26 in Fig. 4) and measure its magnetic response under an external DC magnetic field. In Figs. 6(a) and 6(b) we plot the amplitude and phase of the magnetic response of the magnetometer, measured from the network analyzer, under an external DC magnetic field of $\sim$93~mT. The phase here measures the relative phase between the magnetic response of the magnetometer and the driving field from the network analyzer. Generally, the relative phase changes linearly with frequency. However, there exist abrupt phase changes at the frequencies where the magnetic response exhibits anti-resonance-like dips, as can be seen in the shaded area in Fig. 6(a) and 6(b). In order to study how the amplitude of the DC magnetic field affects the magnetic response of the magnetometer, we gradually decrease the amplitude of the DC magnetic field from 93~mT to 6~mT, with the result shown in Fig. S1(a) in the Supplementary Information. We observe that the magnetic response changes significantly, especially at the frequencies of the dips.


This strong dependence of the magnetic response on the applied DC field observed in Type~II magnetometers can be explained by interference between the magnetostrictive waves generated at each crystal grain, with destructive interference responsible for the observed anti-resonance-like dips in the response. To explore this interference effect, we use a simple multi-wave interference model to simulate the process. Figures 6(c) and 6(d) show the amplitude and phase of the total response from interference of multiple waves. Multiple dips arise naturally in the response spectrum and abrupt phase changes appear at the dip frequencies, due to the destructive interference of different waves. In order to simulate the effect of changing the amplitude of the applied DC magnetic field, we gradually change the relative amplitude of each wave component (simulating modifications of the magnetostriction coefficient due to an external DC magnetic field), with the result shown in Fig. S1(b) in the Supplementary Information. It can be seen that the depth of the dips in the response spectrum changes significantly, e.g. in the shaded area. For more detailed analysis on the study of Type~II magnetic response, see Supplementary Information and Fig. S1.

\subsection{Bandwidth}

\begin{figure}[htb]
\centering
\includegraphics[width=8cm]{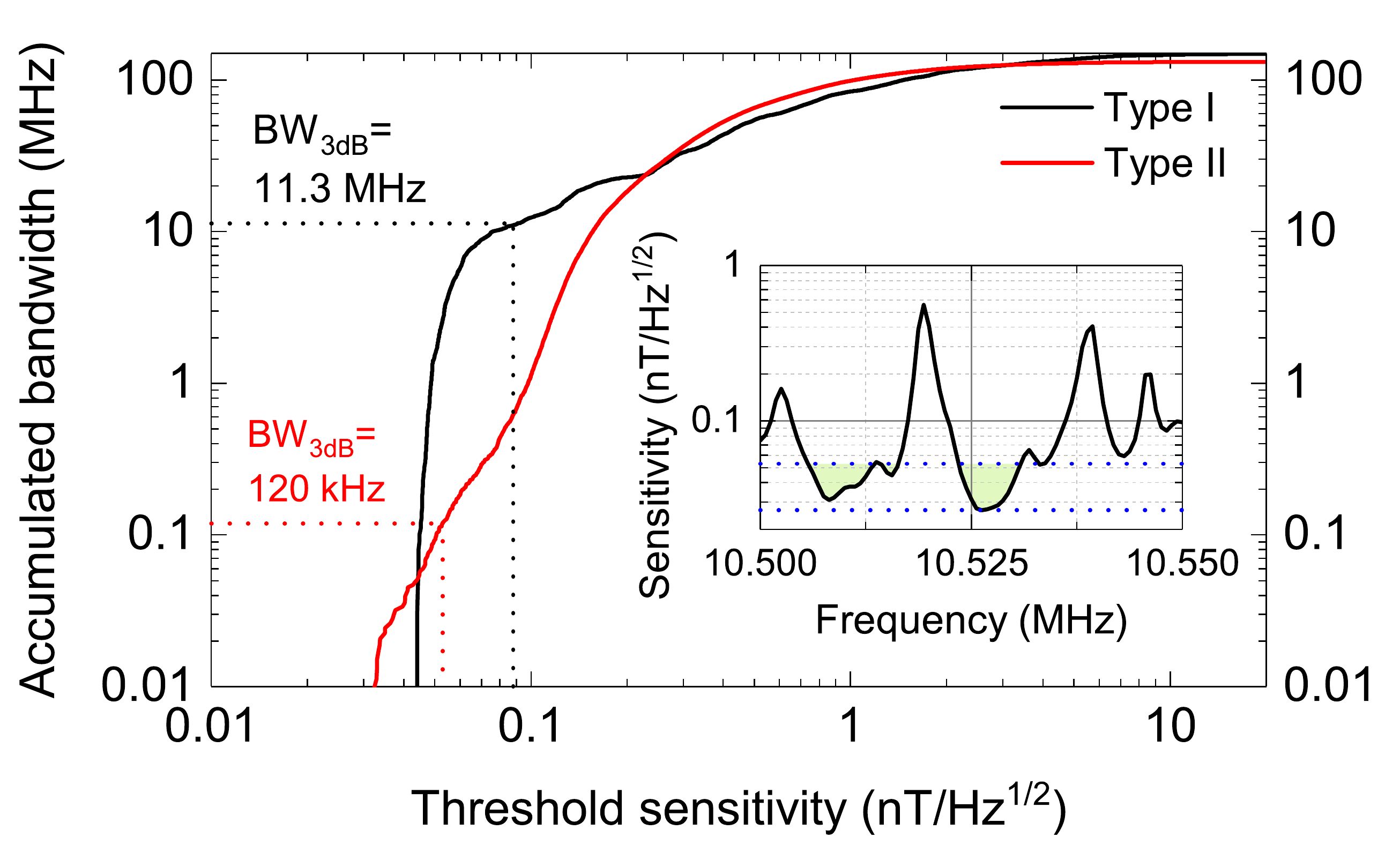}
\caption{(Color online). Accumulated bandwidth as a function of the threshold sensitivity for the Type~I (black curve) and the Type~II (red curve) magnetometers in Fig. \ref{fig2}. The 3~dB bandwidth for the Type~I and Type~II magnetometers are 11.3~MHz and 120~kHz, respectively. In the inset it shows the definition of the accumulated bandwidth, to be the total frequency range in the shaded area.}
\label{fig7}
\end{figure}

We finally discuss the bandwidths of the magnetometers. Due to the fact that the microtoroids support various mechanical modes in a large frequency range from hundreds of kHz to hundreds of MHz, the magnetometers can detect magnetic fields over a broad bandwidth. As can be seen in Figs. 3(c) and 3(f), the working bandwidths for both Type~I and Type~II magnetometers exceed 130~MHz, limited by the bandwidth of the photoreceivers we use in the experiment. As the sensitivity varies with frequency, in order to quantify the bandwidth, we use the accumulated bandwidth \cite{magnetic4} which quantifies the accumulated frequency range over which the sensitivity is better than a certain threshold value. This is illustrated in the inset of Fig. 7, where the accumulated bandwidth is the total frequency range of the shaded area. In Fig. 7, we plot the accumulated bandwidth for a Type~I magnetometer (black curve) and a Type~II magnetometer (red curve), as a function of the threshold sensitivity. The 3~dB bandwidth, over which the sensitivity is within a factor of two of the peak sensitivity, is obtained to be $\sim$11.3~MHz for the Type~I magnetometer, and $\sim$120~kHz for the Type~II magnetometer. This bandwidth is considerably larger than other types of room-temperature magnetometers, such as atomic magnetometers \cite{atomic1,atomic2} and nitrogen vacancy center magnetometers \cite{NV1,NV2}.

\section{Conclusions}

In summary, we have achieved an on-chip, room-temperature magnetometer, with a peak sensitivity of 26~pT/$\sqrt{\mathrm{Hz}}$. This is state-of-the-art for micro-optomechanical magnetometers and is comparable to microscale SQUIDs. We have also found that our magnetometers exhibit two qualitatively different classes of magnetic response. The Type~I magnetometers show a relatively smooth magnetic response as a function of the frequency, following the mechanical resonances of the device; while the magnetic response of the Type~II magnetometers varies significantly over frequency ranges of 10~kHz, within the envelope of the mechanical resonances. We postulate that magnetometers with single crystalline Terfenol-D particles show a Type~I magnetic response, and those with polycrystalline Terfenol-D particles show a Type~II magnetic response, and that the dips in the magnetic response spectra of the Type~II magnetometers arise due to the destructive interference of the magnetostrictive response from different crystal grains. Finally we show that the optomechanical magnetometers can have an ultra-broad accumulated bandwidth. The working bandwidth exceeds 130~MHz for both the Type~I and Type~II magnetometers. The 3~dB bandwidth where the sensitivity is within a factor of two of the peak sensitivity is found to be $\sim$11.3~MHz for a Type~I magnetometer, and $\sim$120~kHz for a Type~II magnetometer, respectively, considerably larger than other sensitive room temperature magnetometers. The high sensitivity and broad bandwidth open up new possibilities for applications such as on-chip microfluidic MRI.

\section*{Acknowledgments}

We thank James Bennett, Varun Prakash, Guangheng Wu, and Enke Liu for the very helpful discussions. This work was primarily funded by DARPA QuASAR Program, Australian Research Council (DP140100734, FT140100650), Australian Defence Science and Technology Group (CERA49 and CERA50), and Commonwealth of Australia as represented by the Defence Science and Technology Group of the Department of Defence. W.P.B. acknowledges an Australian Research Future Fellowship. B.B.L. acknowledges a University of Queensland Postdoctoral Research Fellowship (2014001447). Device fabrication was performed within the Queensland Node of the Australian Nanofabrication Facility.



\begin{thebibliography}{1}

\bibitem{2008Science} T. J. Kippenberg, and K. J. Vahala, "Cavity Optomechanics: Back-Action at the Mesoscale," Science \textbf{321}, 1172-1176 (2008).

\bibitem{2014RMP} M. Aspelmeyer, T. J. Kippenberg, and F. Marquardt, "Cavity optomechanics," Rev. Mod. Phys. \textbf{86}, 1391-1452 (2014).


\bibitem{2014APR} M. Metcalfe, "Applications of cavity optomechanics," Appl. Phys. Rev. \textbf{1}, 031105 (2014).

\bibitem{2008NJP} A. Schliesser, G. Anetsberger, R. Rivi{\`e}re, O. Arcizet, and T. J. Kippenberg, "High-sensitivity monitoring of micromechanical vibration using optical whispering gallery mode resonators," New J. Phys., \textbf{10}, 095015 (2008).




\bibitem{GW} LIGO Scientific Collaboration and Virgo Collaboration, "Observation of Gravitational Waves from a Binary Black Hole Merger," Phys. Rev. Lett. \textbf{116}, 061102 (2016).

\bibitem{force} J. D. Teufel, T. Donner, M. A. Castellanos-Beltran, J. W. Harlow and K. W. Lehnert, "Nanomechanical motion measured with an imprecision below that at the standard quantum limit," Nat. Nanotechnol. \textbf{4}, 820-823 (2009).

\bibitem{mass} W. Yu, W. C. Jiang, Q. Lin, and T. Lu, "Cavity optomechanical spring sensing of single molecules," Nat. Comm. \textbf{7}, 12311 (2016).



\bibitem{acceleration1} A. G. Krause, M. Winger, T. D. Blasius, Q. Lin, and O. Painter, "A high resolution microchip optomechanical accelerometer," Nat. Photonics \textbf{6}, 768-772 (2012).

\bibitem{acceleration2} F. G. Cervantes, L. Kumanchik, J. Pratt, and J. Taylor, "High sensitivity optomechanical reference accelerometer over 10~kHz," Appl. Phys. Lett. \textbf{104}, 221111 (2014).

\bibitem{ultrasound} S. Basiri-Esfahani, A. Armin, S. Forstner, W. P. Bowen, "Precision ultrasound sensing on a chip," Nat. Comm. \textbf{10} 132 (2019).


\bibitem{magnetic1} S. Forstner, S. Prams, J. Knittel, E. D. vanOoijen, J. D. Swaim, G. I. Harris, A. Szorkovszky, W. P. Bowen, H. Rubinsztein-Dunlop, "Cavity Optomechanical Magnetometry," Phys. Rev. Lett. \textbf{108}, 120801 (2012).

\bibitem{magnetic2} S. Forstner, E. Sheridan, J. Knittel, C. L. Humphreys, G. A. Brawley, H. Rubinsztein-Dunlop, W. P. Bowen, "Ultrasensitive Optomechanical Magnetometry," Adv. Mater. \textbf{26}, 6348-6353 (2014).


\bibitem{magnetic3} C. Yu, J. Janousek, E. Sheridan, D. L. McAuslan, H. Rubinsztein-Dunlop, P. K. Lam, Y. Zhang, and W. P. Bowen, "Optomechanical Magnetometry with a Macroscopic Resonator," Phys. Rev. Appl. \textbf{5}, 044007 (2016).


\bibitem{magnetic4} B.-B. Li, J. BÍLEK, Ulrich B. Hoff, L. S. Madsen, S. Forstner, V. Prakash, C. Schafereier, T. Gehring, W. P. Bowen, and U. L. Andersen, "Quantum enhanced optomechanical magnetometry," Optica \textbf{5}, 850-857 (2018).

\bibitem{magnetic5} B.-B. Li, D. Bulla, V. Prakash, S. Forstner, A. Dehghan-Manshadi, H. Rubinsztein-Dunlop,, S. Foster, and W. P. Bowen, "Invited Article: Scalable high-sensitivity optomechanical magnetometers on a chip," APL Photon. \textbf{3}, 120806 (2018).

\bibitem{magnetic6} J. Zhu, G. Zhao, I. Savukov, and L. Yang, "Polymer encapsulated microcavity optomechanical magnetometer," Sci. Rep. \textbf{7}, 8896 (2017).

\bibitem{GHz} M. F. Colombano, G. Arregui, F. Bonell, N. E. Capuj, E. Chavez-Angel, A. Pitanti, S.O. Valenzuela, C. M. Sotomayor-Torres, D. Navarro-Urrios, and M. V. Costache, "Resonant magnon assisted optomechanical magnetometer," arXiv: 1909.03924v1 (2019).


\bibitem{Torque1} J. P. Davis, D. Vick, D. C. Fortin, J. A. J. Burgess, W. K. Hiebert, and M. R. Freeman, "Nanotorsional resonator torque magnetometry," Appl. Phys. Lett. \textbf{96}, 072513 (2010).


\bibitem{Torque2} M. Wu, N. L.-Y. Wu, T. Firdous, F. F. Sani, J. E. Losby, M. R. Freeman, and P. E. Barclay, "Nanocavity optomechanical torque magnetometry and radiofrequency susceptometry," Nat. Nanotech. \textbf{12}, 127-132 (2019).



\bibitem{SQUID1} J. R. Kirtley, M. B. Ketchen, K. G. Stawiasz, J. Z. Sun, W. J. Gallagher, S. H. Blanton, S. J. Wind, "High-resolution scanning SQUID microscope," Appl. Phys. Lett. \textbf{66}, 1138-1140 (1995).

\bibitem{SQUID2} F. Baudenbacher, L. E. Fong, J. R. Holzer, M. Radparvar, "Monolithic low-transition-temperature superconducting magnetometers for high resolution imaging magnetic fields of room temperature samples", Appl. Phys. Lett. \textbf{82}, 3487-3489 (2003).

\bibitem{SQUID3} J. R. Kirtley, L. Paulius, A. J. Rosenberg, J. C. Palmstrom, C. M. Holland, E. M. Spanton, D. Schiessl, C. L. Jermain, J. Gibbons, Y.-K.-K. Fung, M. E. Huber, D. C. Ralph, M. B. Ketchen, G. W. Gibson, and K. A. Moler, "Scanning SQUID susceptometers with submicron spatial resolution," Rev. Sci. Instrum. \textbf{87}, 093702 (2016).



\bibitem{atomic1} H. B. Dang, A. C. Maloof, and M. V. Romalis, "Ultrahigh sensitivity magnetic field and magnetization measurements with an atomic magnetometer," Appl. Phys. Lett. \textbf{97}, 151110 (2010).

\bibitem{atomic2} M. Vengalattore, J. M. Higbie, S. R. Leslie, J. Guzman, L. E. Sadler, and D. M. Stamper-Kurn, "High-resolution magnetometry with a spinor Bose-Einstein Condensate," Phys. Rev. Lett. \textbf{98}, 200801 (2007).

\bibitem{NV1} G. Balasubramanian, P. Neumann, D. Twitchen, M. Markham, R. Kolesov, N. Mizuochi, J. Isoya, J. Achard, J. Beck, J. Tissler, V. Jacques, P. R. Hemmer, F. Jelezko, and J. Wrachtrup. "Ultralong spin coherence time in isotopically engineered diamond," Nat. Matter. \textbf{8}, 383 (2009).

 \bibitem{NV2} T. Wolf, P. Neumann, K. Nakamura, H. Sumiya, T. Ohshima, J. Isoya, and J. Wrachtrup. "Subpicotesla Diamond Magnetometry," Phys. Rev. X \textbf{5}, 041001 (2015).






\bibitem{toroid} D. K. Armani, T. J. Kippenberg, S. M. Spillane, and K. J. Vahala, "Ultra-high-Q toroid microcavity on a chip," Nature \textbf{42}, 925 (2003).

\bibitem{taper} J. C. Knight, G. Cheung, F. Jacques, and T. A. Birks, "Phase-matched excitation of whispering-gallery-mode resonances by a fiber taper," Opt. Lett. \textbf{22}, 1129 (1997).


\bibitem{locking} T. G. McRae, Kwam H. Lee, M. McGovern, D. Gwyther, and W. P. Bowen, "Thermo-optic locking of a semiconductor laser to a microcavity resonance," Opt. Express \textbf{17}, 21977 (2009).













 \bibitem{handbook} G. Engdahl, \textit{Handbook of Giant Magnetostrictive Materials} (Academic Press, 2000).

\bibitem{microstructure1} Y. J. Bi, and J. S. Abell, "Microstructural characterisation of Terfenol-D crystals prepared by the Czochralski technique," J. Cryst. Growth \textbf{172} 440-449, (1997).

\bibitem{microstructure2} G.-H. Wu, X.-G Zhao, J.-H. Wang, J.-Y. Li, K.-C. Jia, and W.-S. Zhan, "<111> oriented and twin-free single crystals of Terfenol-D grown by Czochralski method with cold crucible," Appl. Phys. Lett. \textbf{67}, 2005 (1995).















\end{thebibliography}
\end{document}


\title{Ultra-broadband and sensitive cavity optomechanical magnetometry-Supplementary Information}

\author{Bei-Bei Li}
\affiliation{Institute of Physics, Chinese Academy of Sciences, Beijing 100190, P. R. China.}

\author{George Brawley}
\affiliation{School of Mathematics and Physics, The University of Queensland, St Lucia, Queensland 4072, Australia.}

\author{Hamish Greenall}
\affiliation{School of Mathematics and Physics, The University of Queensland, St Lucia, Queensland 4072, Australia.}

\author{Stefan Forstner}
\affiliation{School of Mathematics and Physics, The University of Queensland, St Lucia, Queensland 4072, Australia.}

\author{Eoin Sheridan}
\affiliation{School of Mathematics and Physics, The University of Queensland, St Lucia, Queensland 4072, Australia.}

\author{Halina Rubinsztein-Dunlop}
\affiliation{School of Mathematics and Physics, The University of Queensland, St Lucia, Queensland 4072, Australia.}

\author{Warwick P. Bowen}
\email{w.bowen@uq.edu.au}
\affiliation{School of Mathematics and Physics, The University of Queensland, St Lucia, Queensland 4072, Australia.}





\date{\today}

\begin{abstract}

In this Supplementary Information, we study the Type~II magnetic response of the magnetometers, including the magnetic response change as a function of the amplitude of the applied DC magnetic field, and a theoretic model that uses multi-wave interference to generate the response similar to the Type~II magnetic response.

\end{abstract}

\maketitle

\section{Type~II magnetic response as a function of a DC magnetic field}

As discussed in the main text, we observed two types of magnetic responses as a function of the frequency of magnetic field, among different magnetometers fabricated using the same method. The Type~I magnetometers exhibit relatively smooth magnetic response as a function of frequency, following the mechanical resonances. The magnetic response of the Type~II magnetometers varies significantly within the frequency ranges of 10~kHz, within the envelope of the mechanical resonances. This Type~II magnetic response has been observed in previous work \cite{magnetic1,magnetic4,magnetic5}, but has not yet been studied. Here in this Supplementary Information, we then study the physical origin of the Type~II magnetic response. As discussed in the main text in Section E, in order to study how the amplitude of the DC magnetic field affects the Type~II magnetic response, we gradually decrease the amplitude of the DC magnetic field from 93~mT to 6~mT, and measure its magnetic response, with the result shown in Fig. S1(a). We can see that the magnetic response changes significantly, especially when close to the dips. For instance, in the shaded area in Fig. S1(a), the magnetic response at $\sim$34.5~MHz changes by more than 25~dB, as the amplitude of the DC magnetic field decreases. This enormous variation in high frequency magnetic response due to the DC field change suggests that DC or low frequency magnetic field sensing can be realized by measuring the change in the magnetic response at high frequencies (in our experimental case, tens of MHz), similar to mixing up the low frequency signals to the high frequency, to improve low frequency magnetic field sensitivity in Ref. \cite{magnetic2}.

\begin{figure*}[htb]
\centering
\includegraphics[width=14cm]{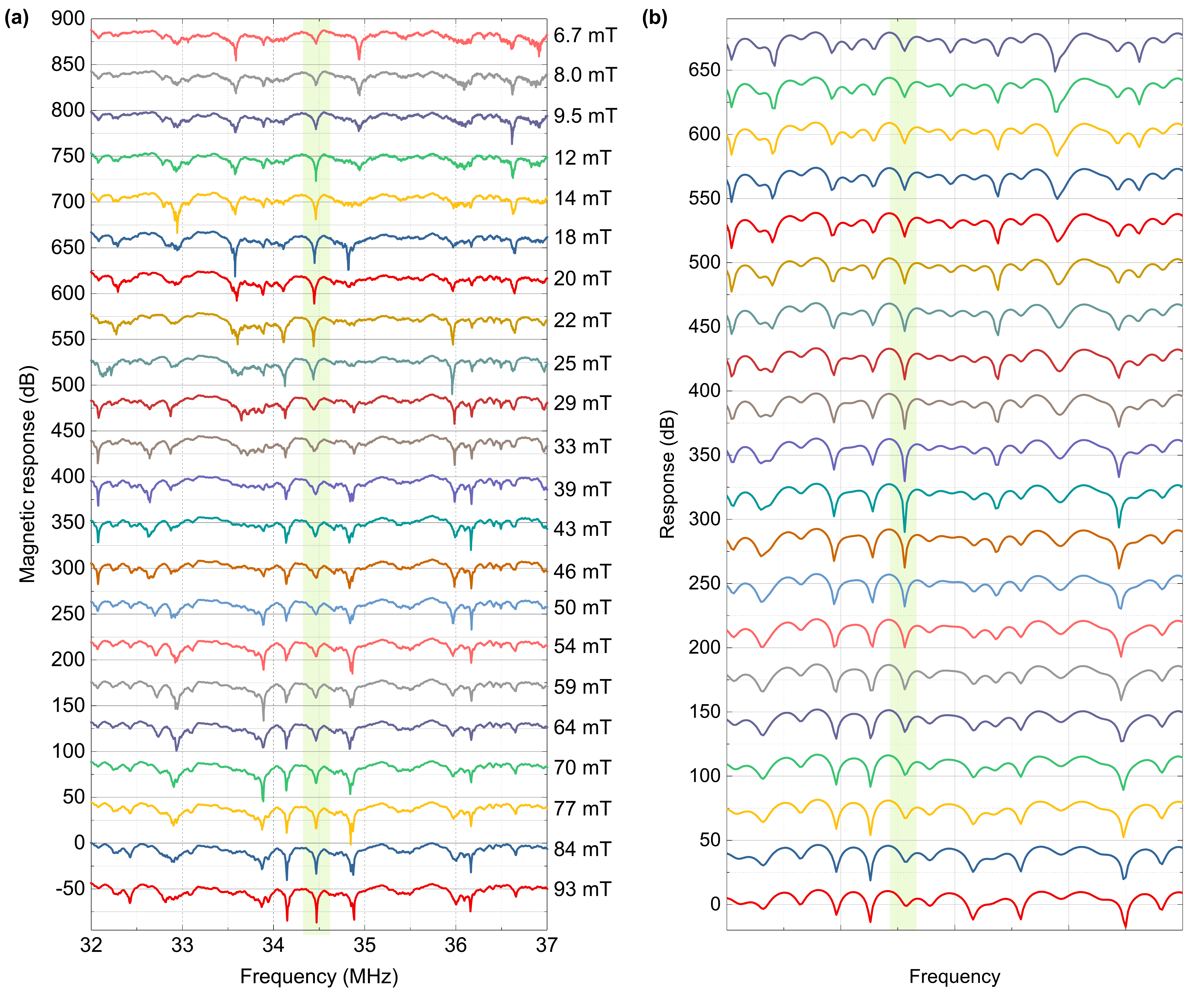}
\caption{(Color online). (a), Magnetic response for different DC magnetic fields applied to the magnetometer, in the frequency range between 32-37 MHz, for a Type~II magnetometer (No. 26 in Fig. 5). (b), Theoretically predicted response obtained from the interference of multiple wave components with different amplitudes and phases, with varied relative amplitudes of the different wave components, which simulates the effect of changing DC magnetic fields applied to the magnetometer.}
\label{figS1}
\end{figure*}

The Terfenol-D particles we use in the experiment are typically around 20~\textmu m to 30~\textmu m in diameter. The commercial Terfenol-D particles are usually manufactured by grinding a larger-sized polycrystal into small pieces \cite{handbook}. As the typical size of the single crystal grain is a few to tens of microns \cite{microstructure1,microstructure2}, the Terfenol-D particles we use for fabricating the magnetometers could be either single crystalline or polycrystalline. We then postulate that the magnetometers with single crystalline Terfenol-D particles exhibit Type~I magnetic response, and those with polycrystalline Terfenol-D particles show Type~II response. In single crystal grains, the crystal plane has a specific angle relative to the external magnetic field, and therefore the magnetostriction is uniform across the whole particle. However, in polycrystalline grains, the crystal planes that are aligned with the external magnetic field can be different in each crystal grain. As the magnetostriction in Terfenol-D crystals is anisotropic, the multiple crystal grains in one Terfenol-D particle will contribute differently to the total magnetostriction. For instance, the magnetostriction coefficient along the [111] direction is the largest, $\lambda_{\mathrm{111}}=1600~ppm$, while that along the [001] direction is much smaller, $\lambda_{\mathrm{100}}=90~ppm$ \cite{handbook}. The magnetic response in the Terfenol-D particle is the total effect of magnetostrictions from different crystal grains. Furthermore, in a magnetostrictive material like Terfenol-D, the application of an external magnetic field induces a stress in the material which results in a strain, i.e., magnetostriction. The relative phase between the strain and the stress can be different in different crystal grains, due to their different intrinsic mechanical resonances. This phase difference causes the magnetostriction out-of-synchronized, especially when Terfenol-D is driven by a high-frequency magnetic field. This means the magnetostrictive waves from different crystal grains have both different amplitudes and phases, and therefore interference needs to be taken into consideration. The magnetostrictions from different grains can produce a destructive interference at some specific frequencies, for instance, at the frequencies where the magnetic response exhibits dips (as shown in Fig. 6(a)), and results in an abrupt phase change at these frequencies (as shown in Fig. 6(b)). When the amplitude of the applied DC magnetic field changes, the magnetostriction from each grain changes due to the domain wall movement in the presence of an external DC magnetic field. This explains why the dip depth in the Type~II magnetic response changes when the amplitude of the external DC magnetic field changes, as shown in Fig. S1(a).





\section{A theoretical model to simulate the Type~II magnetic response}

In order to quantitatively study the interference effect of magnetostriction from different crystal grains, we use a simple model to simulate the process. The total magnetostriction of one Terfenol-D particle is the interference of magnetostrictions from different crystal grains, and each grain contributes a different amplitude and phase to the total magnetostriction. This is very similar to the laser speckle effect, a result of the interference of many waves of the same frequency, having different amplitudes and phases, which add together to give a resultant wave whose amplitude and therefore intensity varies randomly. We therefore use a multi-wave model to simulate the process. The resultant wave from multi-wave interference can be expressed as $A(\omega)=\sum_{j=1}^{N}{A_{j}\exp({i\omega (\Delta T_{j})}})$. Here $A(\omega)$ is the resultant wave at frequency $\omega$ contributed from multiple sources, with $N$ being the number of the sources. $A_{j}$ and $\omega\Delta T_{j}$ are the amplitude and relative phase of the $j$-th source. The number of the sources $N$, representing the number of crystal grains in one Terfenol-D particle, could be random, depending on how that particular Terfenol-D particle is grinded. By assigning random numbers to $A_{j}$ and $\omega\Delta T_{j}$, we can reproduce the total response from multiple sources. In our model, we assume a source number $N=20$, and obtain the total response, with the results shown in Figs. 6(c)-(d) in the main text. Multiple dips arise in the response spectrum (Fig. 6(c) in the main text), due to the destructive interference of different sources, and abrupt phase changes occur at these frequencies (Fig. 6(d) in the main text). These results are very similar to the magnetic response of the Type~II magnetometers. In order to simulate the effect of changing the amplitude of the applied DC magnetic field, we gradually change the relative amplitude of each source (simulating the effect that the magnetostriction coefficient is engineered by an external DC magnetic field), and the total response is shown in Fig. S1(b). It can be seen that the depth of the dips in the spectrum changes significantly e.g., in the shaded area of Fig. S1(b), very similar to that in Fig. S1(a). Similar results are consistently obtained for different $N$ which is varied from a few to a few hundred. This magnetic response measurement not only could be used for DC or low frequency magnetic field sensing by measuring the change in high frequency magnetic response induced by a DC or low frequency magnetic field, but also provides a way to determine whether a Terfenol-D particle is single crystalline or polycrystalline.
